\documentclass[useAMS,usenatbib]{mn2e}
\usepackage{epsf}
\usepackage{psfig}

\def\msun{{h^{-1}{\rm M_\odot}}}
\def\himsun{{h^{-1}{\rm M_\odot}}}
                                             
\def\pppm{\rm P^3M}
\def\mpc{\,h^{-1}{\rm {Mpc}}}

\def\mnras{MNRAS}
\def\apj{ApJ}
\def\apjl{ApJ}
\def\apjs{ApJS}

\def\araa{ARAA}
\def\nat{Nature}
\def\aap{A\&A}

\title[Intrinsic correlation of halo ellipticity and weak lensing
surveys]{Intrinsic correlation of halo ellipticity and its
implications for large-scale weak lensing surveys}

\author[Y.P. Jing]{Y.P. Jing$^{1,2}$\thanks{E-mail:
ypjing@center.shao.ac.cn}\\ $^1$ Shanghai Astronomical Observatory,
the Partner Group of Max-Planck Institut f\"ur Astrophysik, Nandan
Road 80, Shanghai 200030, China\\ $^2$ Beijing Astrophysics Center and
Department of Astronomy, Peking University, 100871 Beijing, China}
\begin{document}
\date{ Accepted ???. Received ??? }
\pagerange{\pageref{firstpage}--\pageref{lastpage}} \pubyear{2002}
\maketitle
\label{firstpage}
\begin{abstract}
We use a large set of state-of-the-art cosmological N-body simulations
[$512^3$ particles] to study the intrinsic ellipticity correlation
functions of halos. With the simulations of different resolutions, we
find that the ellipticity correlations converge once the halos have
more than 160 members. For halos with fewer members, the correlations
are underestimated, and the underestimation amounts to a factor of 2
when the halos have only $20$ particles. After correcting for the
resolution effects, we show that the ellipticity correlations of halos
in the bigger box ($L=300\mpc$) agree very well with those obtained in
the smaller box ($L=100\mpc$). Combining these results from the
different simulation boxes, we present accurate fitting formulae for
the ellipticity correlation function $c_{11}(r)$ and for the
projected correlation functions $\Sigma_{11}(r_p)$ and
$\Sigma_{22}(r_p)$ over three orders of magnitude in halo mass. The
latter two functions are useful for predicting the contribution of the
intrinsic correlations to deep lensing surveys. With reasonable
assumptions for the redshift distribution of galaxies and for the mass
of galaxies, we find that the intrinsic ellipticity correlation can
contribute significantly not only to shallow surveys but also to deep
surveys. Our results indicate that previous similar studies
significantly underestimated this contribution for their limited
simulation resolutions.
\end{abstract} 
\begin{keywords}
Cosmology: theory --- cosmology: observations --- gravitational
lensing --- large-scale structure of universe --- dark matter
\end{keywords}
%
\section{Introduction}
Inhomogeneities of matter distribution in the Universe distort the
images of distant galaxies gravitationally, a phenomenon called
gravitational lensing. The lensing effect induces an
ellipticity-ellipticity correlation of the galaxies on large scales,
which is observable and can be used as a powerful tool to probe the
large-scale dark matter distribution in the Universe (Miralda-Escude
1991; Blandford et al. 1991; Kaiser 1992; see Bartelmann \& Schneider
2001 and Mellier 1999 for reviews).  Several groups have already
detected the ellipticity correlation on scales from $0.4$ to $30$
arc-minutes for faint galaxies (Bacon et al. 2000, 2002; Hoekstra et
al.  2002a,b; Kaiser et al. 2000; Maoli et al. 2001,; Rhodes et
al.~2000, 2001; van Waerbeke et al. 2000, 2001; Wittman et
al. 2000). If the source galaxies are randomly oriented without the
lensing effect, that is, the intrinsic ellipticity correlation of the
galaxies is negligible, the observed ellipticity correlation implies
that the parameter $\beta\equiv \Omega_0^{0.6}\sigma_8$ is about 0.6,
where $\Omega_0$ is the density parameter, $\sigma_8$ is the current
rms linear density fluctuation in a top-hat sphere of $8\mpc$, and $h$
is the Hubble constant in units of $100 {\rm km~s^{-1} Mpc^{-1}}$.

There are however evidences both from theory and from observations
that the shapes of galaxies are intrinsically correlated. In the
theory, the large scale tidal field is expected to induce large-scale
correlations in galaxy spins and in galaxy shapes (Lee \& Pen 2000,
2001; Croft \& Metzler 2000, hereafter CM00; Heavens et al. 2000,
hereafter HRH00; Catelan et al. 2001; Catelan \& Porciani 2001; Hui \&
Zhang 2002; Porciani et al. 2002). It is recently claimed that a
large-scale alignment of galaxy spins has been detected in nearby
galaxy catalogs with a high confidence (Pen et al. 2000, Lee \& Pen
2002; Brown et al. 2002; Plionis 1994 for cluster shapes). While the
intrinsic ellipticity correlation may be separated from the weak
lensing signal in observations through measuring the $E$-mode and the
$B$-mode correlations of the ellipticity (Crittenden et al. 2001,
Mackey et al. 2002), applying this technique needs an accurate
relation between the $E$-mode and $B$-mode correlations (Crittenden et
al. 2001, 2002; Pen et al. 2002; Schneider et al. 2002; Hoekstra et
al. 2002b). In this aspect, the current situations are far from
satisfactory, since there are still considerable uncertainties both in
the theoretical predictions and in the observational measurements of
the intrinsic ellipticity correlations.

Using the numerical simulations of $256^3$ particles released by the
Virgo Consortium, HRH00 and CM00 measured the ellipticity correlation
for dark matter halos and discussed their results in the context of
weak lensing measurements\footnote{HRH00 also examined the correlation
of the halo spins.}.  Assuming that the galaxies have the same
intrinsic correlations as their host halos, they found that the
intrinsic correlation of galaxy ellipticity could dominate over the
lensing signal in shallow lensing surveys, and could contribute a
non-negligible signal, as high as 20 percent, to deep lensing surveys
like those that reported detections of the weak lensing effects. Their
findings already have profound implications for interpreting the weak
lensing experiments, but quantitatively speaking, there are still
significant uncertainties in their results.  For examples, CM00
compared the ellipticity correlation $c_{11}(r)$ for halos of a mass
$>1.4\times 10^{12}\msun$ between two simulations of boxsizes $L=
141\mpc$ and $240\mpc$, and found that $c_{11}(r)$ in the higher
resolution simulation ($L=141\mpc$) is $2\sim 3$ times higher.  The
ellipticity correlation $c_{11}(r)$ found in CM00 is also a factor of
$2$ or more higher than that found in HRH00. Therefore, it is unclear
how their selections of halos and simulation resolutions have affected
their results.

In this {\it Letter}, we will first quantify how the simulation
resolution affects the determination of the ellipticity correlation
using a set of state-of-the-art cosmological N-body simulations of
$512^3$ particles (Jing \& Suto 2002) and a set of lower-resolution
simulations of $256^3$ particles (Jing 1998; Jing \& Suto 1998). We
will show that the minimum particle number $N_{\rm min}$ for resolving
the halo shapes in simulations is $160$, and the ellipticity
correlation functions of halos with more than $N_{\rm min}$ particles
are well converged. We will also show that the ellipticity correlation
is underestimated by a factor $2$ when halos of about twenty particles
are used. Therefore, the predictions of CM00 and HRH00 for weak
lensing surveys are significantly underestimated, since the halos of
ten (HRH00) or twenty (CM00) particles were included in their studies
(in order to have a sufficient number of halos for their
analyses). Our results imply that the intrinsic ellipticity
correlation may have a significant contribution even to deep weak
lensing surveys. Our simulations are also large enough for accurately
measuring the scale- and mass-dependences of the ellipticity
correlation functions. Based on the simulation data, accurate fitting
formulae are presented for these functions. These formulae can be used
to predict the intrinsic ellipticity signals, including $E$-mode and
$B$-mode contributions, in large-scale lensing surveys (Crittenden et
al. 2001, 2002; Schneider et al. 2001; Pen et al. 2002).

In the next section, we will measure the ellipticity correlation
functions in a set of N-body simulations, and will do the convergence
test. In Section 3, we will present the projected ellipticity
correlation functions, which are useful for predicting the intrinsic
ellipticity contribution in weak lensing surveys once the radial (or
redshift) distribution of source galaxies is known. Our results are
summarized and discussed in Section 4.

\section{The ellipticity correlation functions of halos}
\begin{table}
\caption{Cosmological N-body simulations}
 \label{table:cosmosim}
\begin{tabular}{@{}ccccc}
  \hline
$L [\mpc]$ & $N_p$ & $\sigma_8$& $m_p [\himsun]$ & realizations \\
\hline
100 & $512^3$ &0.9 & $6.2\times 10^{8}$ & 4\\
300 & $512^3$ &0.9 & $1.7\times 10^{10}$ & 4\\
100 & $256^3$ &1.0 & $5.0\times 10^{9}$ & 3\\
  \hline
 \end{tabular}
\end{table}

We use a set of cosmological N-body simulations of $512^3$ particles
that are listed in Table 1. The cosmological model is a currently popular
flat low-density model with the density parameter $\Omega_0=0.3$ and
the cosmological constant $\lambda_0=0.7$ (LCDM). The shape parameter
$\Gamma=\Omega_0 h$ and the amplitude $\sigma_8$ of the linear density
power spectrum respectively are 0.2 and 0.9. Two different boxsizes
$L=100\mpc$ and $L=300\mpc$ are adopted, and the particle mass $m_p$
is $6.2\times 10^8\msun$ and $1.7\times 10^{10}\msun$ respectively in
these two cases. Thus, halos of galactic sizes are well resolved in
both types of simulations. The simulation data were generated on the
VPP5000 Fujitsu supercomputer of the National Astronomical Observatory
of Japan with our vectorized-parallel $\pppm$ code (Jing \& Suto
2002), and those of $L=100 \mpc$ have already been used by Jing \&
Suto (2002) to derive a tri-axial model for density profiles of halos
which is significantly more accurate than the conventional spherical
model. There are 4 realizations for each boxsize. For comparison, we
use three realizations of our previous N-body simulations for the LCDM
model that contains $256^3$ particles (Jing \& Suto 1998; Jing
1998). These simulations have the same initial fluctuation phases as
the first three realizations of $512^3$ particles and $L=100\mpc$, but
the normalization $\sigma_8=1.0$ in the lower-resolution simulations
is slightly higher. We refer readers to Jing \& Suto (2002) for
complementary information about the simulations.

Dark matter halos are identified with the friends-of-friends method
(FOF). A linking length $b$ equal to 0.1 of the mean particle
separation is adopted. This linking length is smaller than the nominal
value 0.2. The regions selected $b=0.2$ have a mean density contrast
about 180 (Porciani et al. 2002), which approximately correspond to
virialized halos. The objects identified with $b=0.1$ have a mean
overdensity 8 times higher, and they are actually the inner regions of
virialized halos (i.e. regions within about one third of the virial
radius). Since we are interested in the ellipticity correlation of
galaxies and the galaxies lie in the inner regions of halos,
$b=0.1$ might be more appropriate for our study (see also CM00). The
simulation outputs at redshift $z=1$ are used, for galaxies in deep
lensing surveys are at this redshift. We measure the ellipticity
correlations that are defined (e.g., Miralda-Escude 1991; HRH00; CM00)
as
\begin{equation}
c_{ij}({\bf r})= < e_i({\bf x})e_j({\bf x+r}) >
\end{equation}
where $i$ and $j$ runs over $1$ and $2$, and $\bf r$ is the three
dimensional vector connecting a pair of halos. The ellipticity $e_1$
and $e_2$ for the halos are defined in a projected plane, say, $x$-$y$
plane, and are computed with respect to the axes that are parallel and
orthogonal to the line joining the two halos in the projected
plane. Thus, the ellipticity component $e_1$ ($e_2$) corresponds to
the elongation and compression along (at $45^\circ$ from) the line
joining the two halos in the projected plane. For each sample of
simulations, we compute the correlation functions for the three
projections along the $x$-, $y$- and $z$-axes, and consider these
projections as independent when estimating the errors.

\begin{figure}
\mbox{\psfig{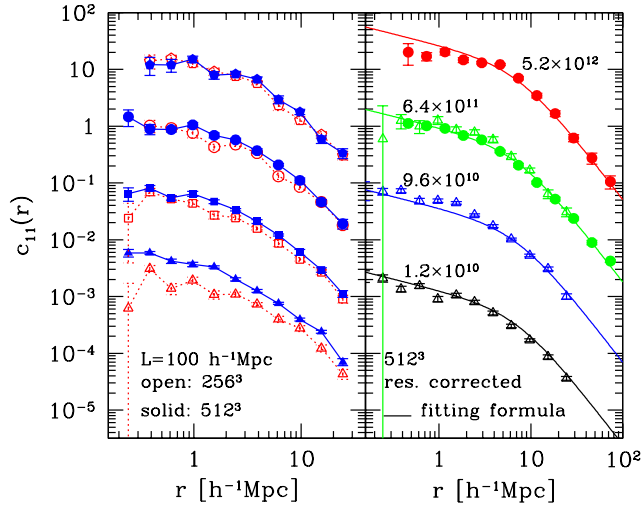}}
\caption{Ellipticity correlation functions of halos $c_{11}(r)$ as a
function of the pair separation $r$. For the clarity, the results for
different halo mass, from bottom to top, are multiplied by a factor of
1, 10, $10^2$ and $10^3$ respectively. {\it left panel}--- The results
are presented for the halos of same spatial number density in the
simulations of $256^3$ particles (open symbols connected with dotted
lines) and in the simulations of $512^3$ particles (solid symbols
connected with solid symbols). From bottom to top, the halos have at
least $20$, $40$, $80$, and $160$ particles respectively in the lower
resolution simulations, and have 7 times more particles in the higher
resolution ones (not $8$ times because $\sigma_8$ is slightly smaller
in the latter). {\it right panels}--- A few typical examples for the
ellipticity correlation functions $c_{11}(r)$ measured in the
simulations of $512^3$ particles in the boxes of $100\mpc$ (open
triangles) and of $300\mpc$ (solid circles). The halos more massive
than $M_h$ are included in the analysis, and the values of $M_h$ (in
$\msun$) are indicated in the figure. The solid line is our fitting
formula (\ref{eq:c11fitting}). The resolution effect has been
corrected according to the left panel.  }
\label{fig:c11}
\end{figure}

Our results confirmed that the cross-correlations $c_{21}({\bf r})$
and $c_{12}({\bf r})$ vanish within the measurement errors (HRH00,
CM00), and these functions will not be discussed anymore. As found in
the previous studies (CM00), we find that the correlation $c_{22}({\bf
r})$ is very anisotropic, but $c_{11}({\bf r})$ is nearly
isotropic. For the presentation convenience, we will discuss the
resolution effect in terms of $c_{11}(r)$.

We analyze $c_{11}(r)$ for halos with particles more than $N_{\rm
min}$ in the simulations of $256^3$ particles and of $512^3$
particles. The simulation box $L$ is $100\mpc$, and only the first
three realizations are used for $N_p=512^3$ (so that the random phases
of the initial fluctuations are the same in the two sets of the
simulations). We consider halos with particles more than $N_{\rm
min}=20$, $40$, $80$, and $160$ in the $N_p=256^3$ simulations. Since
$\sigma_8$ is slightly lower in the $N_p=512^3$ simulations, we fix
the corresponding values of $N_{\rm min}$ for this set of simulations
by requiring that the halos have the same number density $n(>N_{\rm
min})$, and we found $N_{\rm min}=135$, $270$, $540$, and $1080$. The
results are compared in the left panel of Fig.1, which shows that the
correlation function is generally underestimated in the lower
resolution simulations. The underestimation could be caused by an
underestimation of the halo ellipticity and/or a poor determination
of the halo orientations in the lower resolution simulations. We found
that the latter might be the dominant cause, since the mean
ellipticity of halos in the lower resolution simulations are found to be
slightly higher ($<10\%$) than those in the higher resolution
simulations. Nonetheless, the statistical results converge very
rapidly with the resolution of the halos, and $N_{\rm min}=160$ is
sufficient for the simulation measurement. From this test, we found
that the ellipticity correlation is underestimated by a factor of
$2.0$, $1.5$, $1.25$ and $1.05$ for $N_{\rm min}=20$, $40$, $80$, and
$160$. These resolution effects are also confirmed when we compare the
correlation functions from the simulations of different boxsizes with
$N_p=512^3$.

A few examples for $c_{11}(r)$ measured in our simulations of $N_p=512^3$
are presented in the right panel of Fig.~1, illustrating for
halos spanning over 2 orders of magnitude in mass. The
shapes of $c_{11}(r)$ are very similar for different halo mass, and
$c_{11}(r; \ge M_h)$ for halos with a mass greater than $M_h$ can be
well fitted by
\begin{equation}
c_{11}(r; \ge M_h)=\frac{ 3.6\times 10^{-2} 
(\frac{M_h} {10^{10} \msun})^{0.5} }
{r^{0.4}(7.5^{1.7}+r^{1.7})}
\label{eq:c11fitting}
\end{equation}
which are shown in solid lines in the right panel of Fig.1, where $r$
is in units of $\mpc$. The results for the two different boxsizes
agree very well, as shown by the halos of $M_h = 6.4\times
10^{11}\msun$ (the second from top) in the figure.  For the most
massive halos $M_h = 5.2\times 10^{12}\msun$, the correlation function
appears to be slightly flatter than the fitting formula at small
separation $r<1\mpc$, though the discrepancy is small considering the
errorbars of the simulation data. Here $M_h$ is the mass of FOF groups
with $b=0.1$, which is about half of the nominal virial mass of
$b=0.2$ for typical CDM halos.

\section{The projected ellipticity correlation functions}
The angular ellipticity correlation functions $C_{ij}(\theta)$, as
measured in weak lensing surveys, are related to the three dimensional
correlations $e_{ij}({\bf r})$ through a modified Limber's equation
(CM00, HRH00),
\begin{equation}
C_{ij}(\theta)= \frac {\int r_{1}^{2} \phi(r_{1}) r_{2}^{2} \phi(r_{2}) dr_{1} dr_{2}
[1+\xi(r_{12})] c_{ij}(r_p,\pi)}
{\int r_{1}^{2} \phi(r_{1}) r_{2}^{2} \phi(r_{2}) dr_{1} dr_{2}
[1+\xi(r_{12})]}
\label{eq:limber}
\end{equation}
where $r_{12}$ is the comoving separation between the two galaxies at
$r_1$ and $r_2$, and $r_p$ and $\pi$ are the comoving separations
perpendicular to and along the line-of-sight. $\xi(r)$ is the
two-point correlation function, and $\phi(r)$ is the selection
function of a survey. For the flat Universe considered in this paper
and at the small-angle limit (i.e. $\theta\ll 1$ which holds well in
large-scale lensing surveys), we have,
\begin{equation}
r_p=\frac{r_1+r_2}{2}\theta, \hskip 1.5cm \pi=r_1-r_2\,.
\end{equation}
Since the correlations $e_{ij}({\bf r})$ decrease rapidly at scales
$r\ge 10\mpc$ and the selection function $\phi(r)$ changes much more
gently on such scales, equation(\ref{eq:limber}) can simplified to
\begin{equation}
C_{ij}(\theta)= \frac {\int r^{4} \phi^2(r) dr \Sigma_{ij}(r\theta)}
{[\int r^{2} \phi(r) dr]^2+\int r^{4} \phi^2(r) dr \int d\pi
\xi(r\theta,\pi)}
\label{eq:limber_simplified}
\end{equation}
where we define $\Sigma_{ij}(r_p)$ as 
\begin{equation}
\Sigma_{ij}(r_p)\equiv \int d\pi[1+\xi(r_p,\pi)]c_{ij}(r_p,\pi)\,.
\label{eq:pecf}
\end{equation}
which we call the projected ellipticity correlation function. This
function can be measured directly in the simulations.  There are at
least two advantages for adopting $\Sigma_{ij}(r_p)$. First, the
three-dimensional correlations $c_{ij}({\bf r})$ are generally
anisotropic (in fact $c_{22}$ is strongly anisotropic, see CM00), thus
it is easier and more accurate to measure the projected functions.
Second, more importantly, the projected ellipticity correlations
contain all the information needed for predicting the angular
correlation functions $C_{ij}(\theta)$ analytically
(Eq. \ref{eq:limber_simplified}) once the redshift distribution of
source galaxies is known.  .

\begin{figure}
\mbox{\psfig{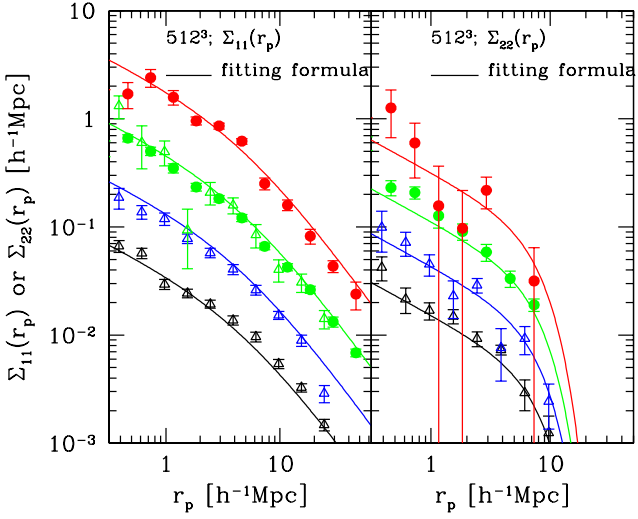}}
\caption{The projected ellipticity correlation functions
$\Sigma_{11}(r_p)$ and $\Sigma_{22}(r_p)$ (defined by
eq.\ref{eq:pecf}) of halos as a function of the project halo
separation $r_p$. The halos are selected and the results are presented
in the same way as in the right panel of Fig.~1, but no vertical shifts
are made for different halos in this plot. The solid curves are given
by fitting formulae (\ref{sig11fitting}) and (\ref{sig22fitting})
respectively for the left and right panels. }
\label{fig:sigma11}
\end{figure}

We plot the projected ellipticity correlation functions
$\Sigma_{11}(r_p)$ and $\Sigma_{22}(r_p)$ measured from our
simulations of $512^3$ particles. The halos have correspondingly the
same mass as those in the right panel of Fig.~1, and the resolution
effects are corrected simply by multiplying the underestimation
factors obtained from the convergence test (the left panel of
Fig.~1). We find that the results from the two simulation boxes agree
very well. The functions $\Sigma_{11}(r_p; \ge M_h)$ and
$\Sigma_{22}(r_p; \ge M_h)$ for halos with mass larger than $M_h$ can
be accurately fitted by the following expressions,
\begin{eqnarray}
\label{sig11fitting}
\Sigma_{11}(r_p; \ge M_h) &=& \frac{0.18(\frac{M_h} {10^{10} \msun})^{0.65}}
{r_p^{0.5}(r_p+5)} \mpc\\
\label{sig22fitting}
\Sigma_{22}(r_p; \ge M_h) &=& 
\frac{1.4\times 10^{-2}(\frac{M_h} {10^{10} \msun})^{0.50}}
{r_p^{0.6}\exp[\frac{1}{2} (\frac{r_p}{6})^{2}]} \mpc
\end{eqnarray}
where $r_p$ is in units of $\mpc$. We note that from the figure, the
readers may get an impression that our fitting formula underestimates
$\Sigma_{22}$ for the most massive halos ($M_h=5.2\times
10^{12}\msun$). But this impression is caused largely by the fact that
the two simulation data points at $r_p=1.2$ and $1.9\mpc$, which are
significantly smaller than the fitting formula, may be mistakingly
considered as those of the less massive halos of $M_h=6.4\times
10^{11}\msun$.

\begin{figure}
\mbox{\psfig{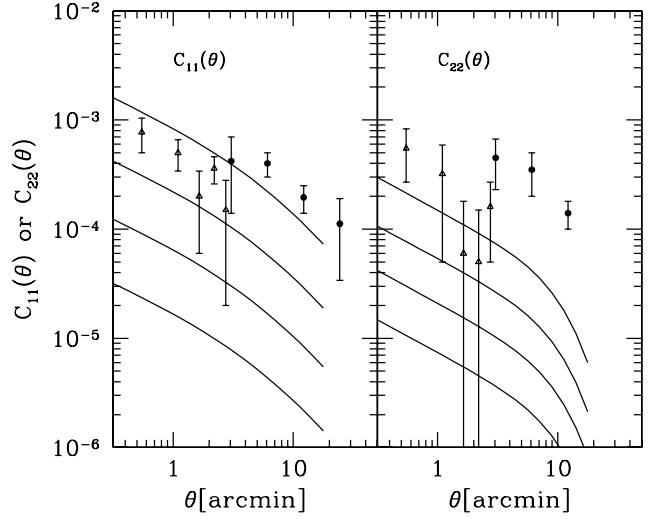}}
\caption{Predictions for the angular ellipticity correlation functions
for halos of mass larger than $ 1.2\times 10^{10}$, $9.6\times
10^{10}$, $6.4\times 10^{11}$, and $5.2\times 10^{12} \msun$ (from
bottom to top), assuming the source redshift distribution has a median
redshift $1$. For comparison, the observational data from deep lensing
surveys of Van Waerbeke et al. (2000; open triangles) and of Wittman
et al. (2000; solid symbols) are plotted.}
\label{fig:c11t}
\end{figure}

Van Waerbeke et al. (2000) and Wittman et al. (2000) have published
their measurements for the angular correlation functions
$C_{11}(\theta)$ and $C_{22}(\theta)$ from their deep surveys.  Our
projected functions can be used to predict the contributions of the
intrinsic correlations to these surveys, if it is assumed that the
galaxies have the same ellipticity as their host halos.  We consider a
redshift distribution function
\begin{equation}
p(z)=\frac{\beta}{\Gamma(\frac{1+\alpha}{\beta})}
(\frac{z}{z_s})^{\alpha}
\exp[-(\frac{z}{z_s})^{\beta}]
\end{equation}
which approximately describes the deep surveys with $\alpha=2$,
$\beta=1.5$ and $z_s\approx 0.7$ (e.g. Smail et al. 1995; Wittman et
al. 2000).  Our predictions for the angular correlation functions,
based on our fitting formulae for $\Sigma_{11}(r_p; \ge M_h)$ and
$\Sigma_{22}(r_p; \ge M_h)$ and the fitting formula of Jing (1998) for
$\xi(r)$ of halos, are shown in Fig.~3, where halo mass $M_h$ from
$10^{10}\msun$ to $5\times 10^{12}\msun$ is considered. If the source
galaxies are more massive than $5\times 10^{11}\msun$, the
contribution of intrinsic ellipticity correlations $C_{11}(\theta)$
becomes comparable to those observed. The intrinsic correlation
$C_{22}(\theta)$ looks much smaller than the observation of Wittman et
al. (2000), but the observed result is also significantly higher than
the predictions of the LCDM model (cf. Wittman et al. 2000) for the
lensing effect. We note that the intrinsic $C_{22}(\theta)$ for $M_h
> 5\times 10^{11}\msun$ is comparable to the prediction of the LCDM
model for the weak lensing.

In contrast to the previous studies of CM00 and HRH00, our results
indicate that the intrinsic correlation of galaxy ellipticity can
contribute significantly not only to shallow surveys but also to deep
surveys. The difference of our conclusions from theirs stems from two
causes. First, CM00 and HRH used the halos with more than $20$ and
$10$ particles respectively to predict the angular correlations, which
can lead to an underestimation of at least a factor of 2 as our
convergence test showed. Second, we show that the ellipticity
correlation increases with the halo mass, and the galaxy mass can be
higher than the mass $2.8\times 10^{11}\msun$ adopted by CM00.

\section{Conclusions}
We used a set of high-resolution cosmological N-body simulations to
study the intrinsic ellipticity correlation functions of halos. With
the simulations of different resolutions, we found that the
ellipticity correlations converge once the halos have more than 160
members. For halos with fewer members, the correlations are
underestimated and the underestimation amounts to a factor of 2 when
the halos have only $20$ particles. After correcting for the
resolution effects, we found that the ellipticity correlations of the
halos in the bigger box ($L=300\mpc$) agree very well with those
obtained in the small box ($L=100\mpc$). Combining these results from
the different simulation boxes, we have presented accurate fitting
formulae for the ellipticity correlation functions $c_{11}(r)$ and for
the projected correlation functions $\Sigma_{11}(r_p)$ and
$\Sigma_{22}(r_p)$ over a large range of halo mass (at least for
$10^{10}\le M_h \le 10^{13}\msun$). The latter two functions can be
used to predict the contribution of the intrinsic correlations to deep
lensing surveys. With reasonable assumptions for the redshift
distribution of galaxies and for the mass of galaxies, we found that
the intrinsic ellipticity correlation can contribute significantly not
only to shallow surveys but also to deep surveys, if the galaxies have
the same shapes as their host halos.

\section*{Acknowledgments}
The author would like to thank Gerhard B\"orner, Joe Silk and Simon
White for useful discussions. The numerical simulations used in this
paper were carried out at ADAC (the Astronomical Data Analysis Center)
of the National Astronomical Observatory, Japan, and Yasushi Suto is
thanked for his support at various stages of obtaining the simulation
sample.  The work is supported in part by the One-Hundred-Talent
Program, by NKBRSF(G19990754), and by NSFC(No. 10125314).

\label{lastpage}
\end{document}